\documentclass[showpacs,preprintnumbers,amsmath]{revtex4}
\usepackage{graphicx}

\begin{document}
\title{Stochastic Resonance in an Overdamped Monostable System.}
\author{Nikolay Agudov and Alexey Krichigin}
\email{agudov@rf.unn.ru} \affiliation{ Faculty of Radiophysics, University of Nizhny Novgorod, 23 Gagarin Ave, Nizhny Novgorod, 603950, Russia}
\date{August 14, 2008}

\begin{abstract}
We show, the SR can appear in monostable overdamped systems driven
by additive mix of periodical signal and white Gaussian noise. It
can be observed as non-monotonic dependence of SNR on the input
noise intensity. In this sense it is similar to classical SR
observed in overdamped bistable systems with potential barrier.
\end{abstract}

\maketitle

The dynamics of nonlinear periodically driven stochastic systems
has been attracted great attention during the last decades . The
interest in these systems is much stimulated by phenomenon known
as stochastic resonance (SR), where noise plays a constructive
role \cite{Anishchenko02,Gammaitoni98}. This phenomenon can be
defined as enhancement of sensitivity of a nonlinear system to
external periodical forcing. Nowadays the SR has been found and
studied in a different physical, chemical and biological systems
\cite{Dykman95,Greenwood00,Guo06,Mitaim98,Vilar99,Jung02,Zozor03,Valenti04,Spagnolo02}.
The enhancement of sensitivity is usually understood as
nonmonotonic dependence of the signal-to-noise ratio (SNR) or
signal power amplification (SPA) at the output of nonlinear system
as a functions of the input noise intensity. Accordingly, the
phenomenon of SR displays itself, when SNR and SPA reach maximum
at some value of noise intensity and then decrease with further
growth of fluctuations.

The effect of SR was observed in various monostable systems: with
multiplicative noise \cite{Guo06}, signal array \cite{Lindner01},
underdamped \cite{Stocks93}, higher harmonics \cite{Grigorenko97}.
The present paper is dedicated to the investigation of monostable
systems described by the overdamped Langevin equation  with an
additive noise and an additive driving signal
\begin{equation}\label{lan}
{{dx}\over{dt}}=-{{d\Phi(x)}\over{dx}}+s(t)+\xi(t),
\end{equation}
where $s(t)=A\cos(\omega_0t)$ is the input driving signal,
$\xi(t)$ is the input white Gaussian noise: $<\xi(t)>=0$,
$<\xi(t)\xi(t+\tau)>=2q\delta(\tau)$, $2q$ is the noise intensity,
$\Phi(x)$ is potential field describing the system and $x(t)$ is
the output random process.

The canonical example of SR was observed and studied in overdamped
system (\ref{lan}) for the bistable potential profile $\Phi(x)$
with single potential barrier separating the metastable states in
Refs.\cite{Anishchenko02,Gammaitoni98,McNamara89}.  This result
can be generalized for multistable potentials with arbitrary
number of barriers. The value of additive noise intensity for
which SNR reaches the maximum was revealed to be about the height
of the potential barrier. In other words, the presence of
potential barrier(s) has been considered as necessary condition
for arising of SR in overdamped systems with additive noise
(\ref{lan}). It is well known also that in bistable (multistable)
systems the non-monotonic dependence of SNR is accompanied by the
similar non-monotonic dependence of SPA on noise intensity.

Recently, in Ref.~\cite{Evstigneev04} it was shown that some
special kind of SR can appear in overdamped monostable systems
(\ref{lan}), where there is no any barriers in potential profile
$\Phi(x)$. This kind of SR is different from classical one,
because the non-monotonic dependence on the noise intensity is
observed only for SPA. While the SNR was shown to be monotonically
decreasing function of the  noise intensity  regardless to various
non-monotonic dependencies of SPA \cite{Agudov08}.

In the present Letter we demonstrate that the non-monotonic
dependence of SNR, similar to that for bistable systems, can be
observed also in monostable overdamped systems. This result
implies the presence of SR in monostable overdamped system with
additive noise and signal. For this case the maximum of SNR for
non-zero noise level has never been observed before. On the other
hand, this situation is sufficiently general to be achieved in a
great vaiety of physical, chemical, and biological systems.

The function of SNR is obtained analytically. The analysis of the
SNR and SPA shows that this phenomenon of SR is new and it has
different properties comparing to the classical case appearing in
the systems with barriers.

Consider the following piece-wise linear monostable potential
(See Fig.~1)
\begin{equation}\label{pot}
\Phi(x)=\left\{
\begin{array}{ll}
k_1 |x|, & |x|<L,\\
k_2 (|x|-L)+k_1 L, & |x|>L.
\end{array}
\right.
\end{equation}
\begin{figure}[htpb]
\begin{center}
\label{Fig_01} \centering{\rule{ 0cm}{0cm}}
\includegraphics[width=60mm]{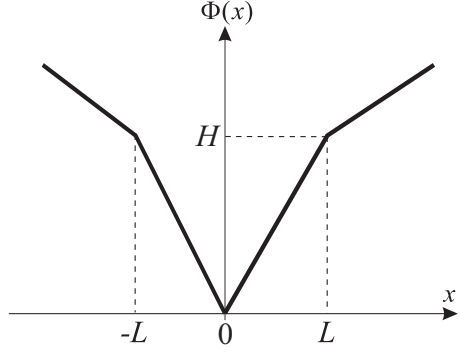}
\caption{Piece-wise linear monostable potential $\Phi(x)$.}
\end{center}
\end{figure}

This potential profile is monostable and has two parameters
specifying the slope of potential profile wells: $k_1$ describes
the slope near the minimum  at $|x|<L$ and $k_2$ for $|x|>L$. The
both values $k_1$ and $k_2$ are always positive providing the
monostability (single minimum) of potential $\Phi(x)$, while $k_1$
can be grater than $k_2$ and vice versa.

For derivation of the power spectrum density (PSD) of the output
signal $x(t)$ the linear response theory (LRT) is used assuming
that the magnitude $A$ of driving signal $s(t)$ is small enough:
$A\ll \min\{k_1,k_2\}$. In accordance with LRT (See for example
Ref.~\cite{Anishchenko02}) the PSD of the output process $x(t)$
reads
\begin{equation}\label{SpectrLRT}
S_x(\omega)=S_x^{(0)}(\omega)+{{a^2}\over{4}}\left(\delta(\omega-\omega_0)+\delta(\omega+\omega_0)\right).
\end{equation}
The function $S^{(0)}_x(\omega)$ provides the noise platform and
the other term is the output signal with amplitude $a$. Therefore
the SNR is defined as follows \cite{Anishchenko02}
\begin{equation}\label{SNR}
R={{a^2}\over{2 S_x^{(0)}(\omega_0)}}.
\end{equation}

The function $S^{(0)}_x(\omega)$ is the PSD of the unperturbed
system (\ref{lan}) under $s(t)=0$, which is defined as the Fourier
transform of the appropriate unperturbed autocorrelation function
\begin{equation}\label{Spectr}
S_x^{(0)}(\omega)={1\over\pi} \int\limits_{0}^{\infty}
K_x^{(0)}[\tau] cos(\omega\tau) d\tau,
\end{equation}
where $K_x^{(0)}[\tau]=\left<x(t) x(t+\tau)\right>$. In the above
expression we have taken into account that $K_x^{(0)}[\tau]$ is
even function. According to the LRT, the amplitude of output
signal is
\begin{equation}
\label{OutputSignal} a=A|\chi(i\omega_0)|,
\end{equation}
where $\chi(i\omega)$ is the susceptibility of the system.
Therefore SPA of input signal $s(t)$ reads
\begin{equation}\label{SPA}
\eta={a^2\over A^2}=|\chi\left(i\omega_0\right)|^2.
\end{equation}
The susceptibility is the Fourier transform of linear response
function $h(\tau)$
\begin{equation}\label{chi}
\chi(i\omega)=\int\limits_{-\infty}^{\infty}h(\tau)e^{-i\omega\tau}d\tau,
\end{equation}
while the linear response function can be expressed in terms of
correlation function of unperturbed system in accordance with
fluctuation-dissipation theorem
\begin{equation}
\label{FDT3}
{h(\tau)}=-{{\theta(\tau)}\over{q}}{{dK_x^{(0)}[\tau]}\over{d\tau}},
\end{equation}
where $\theta(\tau)$ is the Heaviside function.

The probability density function (PDF) of the unperturbed process
$W^{(0)}(x,t)$ satisfies the Fokker-Planck equation (FPE)
\cite{Risken89}
\begin{eqnarray} \label{FPE}
{{\partial{W^{(0)}(x,t)}}\over{\partial{t}}} =
{\partial\over\partial
x} \left({d\Phi(x)\over{dx}}W^{(0)}(x,t)+ \right. \\
\nonumber \left. +q{\partial{W^{(0)}(x,t)}\over\partial{x}}\right)
\end{eqnarray}
with boundary conditions $W^{(0)}(\pm\infty, t)=0$. Since we
consider $\Phi(x)\to\infty$  under $x\to\pm\infty$, the stationary
PDF will be established in the system with time
\begin{equation}
\label{Wst} W_{st}^{(0)}(x)=N \exp\left(-{\Phi(x)\over q}\right),
\end{equation}
where $N$ is the normalization factor. Therefore to find
autocorrelation function
\begin{equation}\label{KorFun}
K_x^{(0)}[\tau]=\int\limits_{-\infty}^{\infty}x_0W_{st}^{(0)}(x_0)dx_0\int\limits_{-\infty}^{\infty}xW^{(0)}(x_0|x,\tau)dx.
\end{equation}
it is necessary to obtain the transition probability density
$W^{(0)}(x_0|x,t)$, which is the solution of FPE (\ref{FPE}) with
the initial conditions $W^{(0)}(x,0)=\delta(x-x_0)$.

For the real physical system the integration in Eq.~(\ref{chi})
can be not from $-\infty$ but from $0$, because linear response
function, according to Eq.~(\ref{FDT3}), exists only if $\tau >
0$. Therefore susceptibility $\chi(i\omega)$ in Eq.~(\ref{chi})
can be considered as the Laplace transform of the linear response
function
\begin{equation}\nonumber
\chi(p)= \hat{h}(p)=\int_0^\infty h(\tau)e^{-p\tau}d\tau,
\end{equation}
where $p=i\omega$ is the Laplace variable. On the other hand, we
can find Laplace transform of the linear response function by
integrating the expression (\ref{FDT3}) and finally we obtain
\begin{equation}\label{chi_lap}
\chi(i\omega)={1\over{q} } \left( K_x^{(0)}[0] -
i\omega\hat{K}_x^{(0)}[i\omega] \right),
\end{equation}
where $K_x^{(0)}[0]$ is the correlation function at $\tau=0$
\begin{equation}\label{Kor_0}
K_x^{(0)}[0]=\langle x^2\rangle+m^2_{st},
\end{equation}
which is expressed in terms of variance and mean value of the
stationary distribution (\ref{Wst}). The PSD (\ref{Spectr}) also
can be written as the real part of the Laplace transform $\hat
K_x^{(0)}[p]$
\begin{equation}\label{LSpectr}
S_x^{(0)}(\omega)={1\over\pi} \mathbf{Re} \left\{\hat K_x^{(0)}[i\omega]\right\}.
\end{equation}

In the present paper the Laplace transform of autocorrelation
function is obtained for monostable potential (\ref{pot}). The
Laplace transform method for solution of the FPE is described in
Refs.~\cite{Atkinson68,Agudov93,Malakhov96,Privman91}. In
particular, in Ref.~\cite{Agudov93} the exact Laplace transform of
transition probability density is obtained for piece-wise linear
potential profile consisting of an arbitrary number of linear
parts. Using this approach we obtain the following exact
expression for Laplace transform of unperturbed autocorrelation
function
\begin{equation}\label{KorFunLapEnd}
\hat{K}_x^{(0)}[p]= {{\langle
x^2\rangle}\over{p}}+{{\sum\limits_{i=0}^{9}A_i}\over{B}},
\end{equation}
where coefficients $A_i$ are as follows
\begin{equation}\nonumber
{\langle
x^2\rangle}={{q}\over{2p\beta\delta^2}}{{2+2h\delta+h^2\delta^2-\left(h^2+2h+2\lambda\right)\delta^3}\over{1-\lambda\delta}},
\end{equation}
\begin{equation}\nonumber
A_0={{4q\beta\lambda\mu\left(1-4\delta\right)}\over{\left(1-\lambda\delta\right)}},
\end{equation}
\begin{equation}\nonumber
A_1=-{{8q\beta\left(1-\delta\right)\left(\left(1-\lambda\right)\alpha_1^2+\alpha_2^2\right)\gamma_1}\over{\left(1-\lambda\delta\right)}},
\end{equation}
\begin{equation}\nonumber
A_2={{2q\left(2\lambda\mu\left(2-\delta\right)-4\delta\mu+\delta\lambda\left(2-\mu\right)\gamma_1\right)}\over{\left(1-\lambda\delta\right)}},
\end{equation}
\begin{equation}\nonumber
A_3=-{{4q\gamma_1\left(\left(2-\delta\right)\alpha_2^2+\delta\lambda\left(2+\left(2-\mu\right)\gamma_2\right)\right)}\over{\delta^2\left(1-\lambda\delta\right)\left(1+\gamma_2\right)}},
\end{equation}
\begin{equation}\nonumber
A_4=-{{4q\gamma_1\alpha_1^2\left(2-3\delta\right)\left(1-\lambda\right)}\over{\delta^2\left(1-\lambda\delta\right)\left(1+\gamma_2\right)}},
\end{equation}
\begin{equation}\nonumber
A_5={{2q\mu\left(\left(2-\delta\right)\lambda+2\delta\left(2-\lambda\right)\gamma_2+2\delta\left(1-\lambda\right)\right)}\over{\delta^2\left(1-\lambda\delta\right)\left(1+\gamma_2\right)}},
\end{equation}
\begin{equation}\nonumber
A_6={{q\gamma_1\left(2-\mu\right)\left(\delta\lambda\left(1+\gamma_2\right)-6+2\delta-2\gamma_2\right)}\over{\beta\delta^2\left(1-\lambda\delta\right)\left(1+\gamma_2\right)}},
\end{equation}
\begin{equation}\nonumber
A_7={{2q\mu\left(1+\delta\left(2-\lambda\right)-2\delta^2\lambda\right)\gamma_2}\over{\beta\delta^2\left(1-\lambda\delta\right)\left(1+\gamma_2\right)}},
\end{equation}
\begin{equation}\nonumber
A_8={{2q\mu\left(3+4\lambda-2\delta^2\lambda-\delta\left(2+3\lambda\right)\right)}\over{\beta\delta^2\left(1-\lambda\delta\right)\left(1+\gamma_2\right)}},
\end{equation}
\begin{equation}\nonumber
A_9={{4q\mu}\over{\beta^2\delta^3\left(1+\gamma_2\right)}},
\end{equation}
\begin{equation}\nonumber
B=p^2\left(1+\gamma_2\right)^2\left(\left(2-\mu\right)\gamma_2-\mu\left(1-\delta+\delta\gamma_2\right)\right),
\end{equation}
here $h={{k_1L}/{q}}$, $\beta={{k_2^2}/{2pq}}$,
$\gamma_1=\sqrt{1+{{4pq}/{k_1^2}}}$,
$\gamma_2=\sqrt{1+{{4pq}/{k_2^2}}}$, $\delta={{k_2}/{k_1}}$,
$\lambda=1-e^h$, $\mu=1-e^{h\gamma_1}$,
$\alpha_1=\sqrt{1-\mu}-{{1}/{\sqrt{1-\lambda}}}$,
$\alpha_2=\sqrt{1-\mu}-\sqrt{1-\lambda}$.

\begin{figure}[htpb]
\begin{center}
\label{Fig_02} \centering{\rule{ 0cm}{0cm}}
\includegraphics[width=60mm]{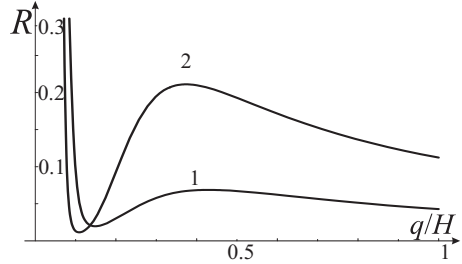}
\caption{The SNR of the system (\ref{lan}) with monostable
potential (\ref{pot}) versus dimensionless noise intensity $q/H$
for the amplitude of input signal $A=0.3$ and frequency
$\omega_0=0.1$. The Curve 1 corresponds to $k_1=25$, $k_2=3$,
$L=0.2$ and Curve 2 corresponds to $k_1=20$, $k_2=1$, and
$L=0.1$.}
\end{center}
\end{figure}
With the above exact expression for autocorrelation function
(\ref{KorFunLapEnd}) we can obtain the PSD (\ref{LSpectr}) and the
SNR (\ref{SNR}). In Fig.~2 the SNR is plotted versus the
dimensionless input noise intensity $q/H$ for different values of
$L$ and the slopes $k_1$, $k_2$. Namely, curve 1 is for $L=0.2$,
$k_1=25$ and $k_2=3$, curve 2 is for $L=0.1$ and $k_1=20$ and
$k_2=1$. The value $H=k_1L$ is the parameter of the potential
profile (\ref{pot}) (See the Fig.~1). As one can see from  Fig.~2,
for small and large $q$, the SNR is decreasing function of $q$
similar to the other cases of monostable potentials considered
earlier in Ref.~\cite{Agudov08}. While for the intermediate values
of noise intensity $0.1H\leq q\leq0.4H$ the non-monotonic behavior
of the SNR appears. It looks similar to the effect of the SR
observed in bistable systems with barrier (See for example
Refs.~\cite{Anishchenko02, Gammaitoni98}).

The properties of this SR behavior are varying, depending on the
values of parameters of the investigated monostable potential
profile (\ref{pot}). The effect is observed when $k_1>3k_2$. It
follows from Fig.~2, that the maximum of SNR is reached at noise
intensity $q^*\simeq 0.4H$. The difference between the values of
SNR in the maximum and in the minimum is growing with the
difference between $k_1$ and $k_2$. In particular, for the Curve 2
in the Fig.~2 we can see the improvement in SNR about 10 times
with increasing of input noise from $q=0.1H$ up to $q^*=0.4H$.

In spite of the similar manifestation, the mechanism of this SR
should be different from that in bistable systems, where the key
parameter of the SR effect is the height of potential barrier. In
the considered monostable system there is no barrier and the
force, which is regular in time, always tends to return the system
to the equilibrium point $x=0$, corresponding to the minimum of
potential profile (\ref{pot}). The SR appears for the special
shape of potential profile, which defines the strength of the
regular force. Namely, the regular returning force should be much
weaker in the area located far from equilibrium point comparing to
that being near the equilibrium. For the potential (\ref{pot}) the
strength of the force is changed in the points $x=\pm L$. If this
change is large enough $k_2<3k_1$, the SR can be observed as a
maximum of SNR as a function of the noise intensity.

Such a shape of potential profile provides the system properties,
which are similar to those for excitable systems, where the SR
also was observed \cite{Gammaitoni98,Longtin93,Longtin98}. The
excitable systems also have only one stable state. Under a small
perturbation these systems relax quickly to the stable state.
While a large (over threshold) perturbation switches system to an
excited state. The exited state is not stable and it decays to the
stable state but after relatively longer time. For the potential
profile (\ref{pot}) when $k_1\gg k_2$ the system (\ref{lan})
returns to the stable state relatively quickly, if a perturbation
is less than $L$. When a perturbation exceeds the threshold value
$L$, the system delays in the region $|x|>L$ for a relatively
longer time, because the returning force there is weak. On the
other hand, the similarities between the investigated and
excitable systems are only qualitative. The equations describing
excitable and threshold systems are different from
Eq.~(\ref{lan}).

\begin{figure}[htpb]
\begin{center}
\label{Fig_03} \centering{\rule{ 0cm}{0cm}}
\includegraphics[width=60mm]{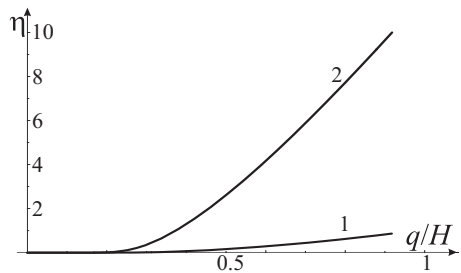}
\caption{The SPA of the system (\ref{lan}) with monostable
potential (\ref{pot}) versus dimensionless noise intensity $q/H$
for the amplitude of input signal $A=0.3$ and frequency
$\omega_0=0.1$. The Curve 1 corresponds to $k_1=25$, $k_2=3$,
$L=0.2$ and Curve 2 corresponds to $k_1=20$, $k_2=1$, and
$L=0.1$.}
\end{center}
\end{figure}

The properties of the SR effect observed here for the monostable
potential (\ref{pot}) has another important difference from the SR
in bistable systems. In accordance with analysis carried out by
various authors, in bistable (and multi-stable) systems the
non-monotonic behavior of the SNR with the input noise is
accompanied  by similar behavior of the SPA. Using the Laplace
transform of the autocorrelation function (\ref{KorFunLapEnd}), we
can obtain the SPA for the investigated monostable system. The
plot of SPA as a function of the input noise intensity is shown in
Fig.~3. The parameters of the system for the Curves 1 and 2 in the
Fig.~3 are the same as those in Fig.~2, where the plots of SNR are
presented. One can see, the non-monotonic dependence of SNR
corresponds to the monotonically growing behavior of SPA as a
function of noise intensity.

This diversity in the properties for the SR in bistable and
monostable systems confirms the assumption about different
mechanisms for the SR effect in these systems. Therefore we can
conclude that the SR effect presented here is new and needs
further detailed analysis.

The authors acknowledge fruitful discussions with Bernardo Spagnolo.
This work is supported by Russian Foundation for Basic Research
(project 08-02-01259).

\end{document}